# Quantum corrections to gravitational potential of scalarized neutron star binary

J. Wang[a]

School of Physical science and Technology, Guangxi Normal University, Guilin 541004, People's Republic of China

**Abstract** We investigate the long-distance, low-energy, leading quantum corrections to gravitational potential for scalarized neutron star (NS) binary systems, by treating general relativity as an effective field theory. We neglect the extended scales of two star components and treat them as heavy point particles, which gravitationally interact with each other via the exchanges of both gravitons and scalar particles, because of the settled scalar configurations inside the stars. Accordingly, the gravitational potential includes both Newtonian potential and scalar-modified Newtonian-like part. We, in the non-relativistic limit, calculate the non-analytic corrections to the modified gravitational potential directly from the sum of all exchanges of both gravitons and scalar particles to one-loop order. The appropriate vertex rules are extracted from the effective Lagrangian. Our calculations demonstrate that either the graviton exchanges or the exchanges of scalar particles contribute to both classical relativistic corrections and quantum corrections to the gravitational potential of the scalarized NS binaries.

## 1 Introduction

According to classical gravitational theory, the Newtonian potential

$$V_{\text{Newt}} = -\frac{Gm_1 m_2}{r} \tag{1}$$

is approximately valid for the gravitational interactions in compact binary,[1] such as neutron star (NS) binary. With a definition of bound state potential in the framework of general relativity, the relativistic corrections, arising from higher order effects in $\frac{v^2}{c^2}$ and nonlinear terms in the field equations of order $\frac{Gm}{c^2 r}$ ($m = m_1, m_2$), in a Hamiltonian treatment were completed [1], which was obtained in the form of

$$V_{\text{cl}} = a_{\text{cl}} \frac{Gm_1 m_2}{r} \frac{G(m_1 + m_2)}{c^2 r}, \tag{2}$$

where $a_{\text{cl}}$ is a numerical constant that would depend on the precise definition of the potential. The classical relativistic corrections to the interaction potential of two bodies also were discussed [2,3], with general agreement with the above result, although in unavoidably ambiguously defining the potential.

The theory of general relativity has been widely accepted in describing the gravitationally bound systems consisting of two compact extended objects. The post-Newtonian approximation to general relativity [4,5], i.e. systematically solving the Einstein equations with nonrelativistic sources, is employed as the conventional approach to calculating the initial inspiral of the system as it slowly loses energy to gravitational radiation [6]. In order to investigate the gravitational radiation power spectra emitted by nonrelativistic bound systems, the dynamics of two components, which are treated as point particles, in a binary system coupled to gravity was described in an effective field theory framework [7], in which the observables appearing in long-wavelength physics are consistent with general coordinate invariance of general relativity.

General relativity, with low energy degree of freedoms and gravitational interactions, is a consistent effective field theory [8], which allows, in principle, its quantization to be carried out without knowledge of microphysics details. By using the effective field theory approach with background field quantization [9,10], the leading long-distance quantum corrections to the one-particle-irreducible potential were calculated [8,11], which result in a finite correction,

$$V_{\text{qu}} = a_{\text{qu}} \frac{Gm_1 m_2}{r} \frac{G\hbar}{c^3 r^2}, \tag{3}$$

---

[1] $G$ is the Newtonian gravitational constant. $m_1$ and $m_2$ are the masses of two components in the binary system.

[a] e-mail: joanwangj@mailbox.gxnu.edu.cn

Springer

where $a_{qu}$ is a numerical constant. However, many works dedicated to the choices between various definitions of the potential depending on the physical situation and the way of defining the total energy. By using the Arnowitt-Deser-Misner formula for the total energy of the gravitational system, the Wilson loop description for the gravitational potential has been done [12,13]. The quantum corrections to Newtonian potential for an arbitrary gravitational field that includes the back-reaction produced by a quantum scalar field of mass was considered by deriving an in-in effective equations [14]. For the simplicity and intuitiveness, a number of authors employed the scattering amplitude itself to define the potential [2,15–19]. The obtained particular effects, from summing one-loop Feynman diagrams with off-shell gravitons, applies to point particle masses [19]. The quantum gravitational effects of a pair of localized polarizable objects, associated with two-graviton exchange from the induced gravitational quadrupole moments due to quantum fluctuations in the metric, was computed [20].

Several observations of binary pulsar systems, i.e. Hulse–Taylor system, PSR 1913+16 [21,22], and PSR J1738+0333 [23] indicated an excess orbital decay, which directly translates to a dipole radiation constraint on the deviations from the quadruple formula. It was relived by considering that a nontrivial scalar configuration comes about in strong-field regime [24]. In analogy with the spontaneous magnetization of ferromagnets below the Curie temperature, an NS, with a compactness above a certain critical value, will occurs spontaneous scalarization [25]. NSs, with a mass of 1.4 $M_\odot$, in binary system would develop strong scalar charges even in absence of external scalar solicitation for strong couplings [26], which enhances the gravitational interactions with the companion star and induces its scalarzation [27]. The iterative interplay between two scalarized components generates a gravitational scalar counterpart, besides the gravitational tensor radiation, which was assigned to a dynamical scalarization of the system [28]. As a consequence, the gravitational interactions of two scalarized components in a scalarized NS binary include exchanges of both gravitons and scalar particles. In this work, we investigate the NS binary systems with orbital separations of about $10^9$ m, which is expect to coalesce and merge in the Hubble time. Consequently, the relative momentum between two star components can be neglected, and we just consider the velocity-free gravitational potential, which depends on only the orbital separation of the system. Accordingly, the Newtonian gravitational interaction is modified by the gravitational scalar interactions, which reads [25]

$$V_{\text{scal}} = -G\frac{m_1 m_2}{r} - G\frac{\omega_1 \omega_2}{r}, \quad (4)$$

where $\omega_1$ and $\omega_2$ denote the scalar charges developed inside the two scalarized components [29].

The exchanges of both gravitons and scalar particles involve scattering process, i.e. the gravitational scattering of two scalar particles, the gravitational scattering of gravitons by scalar particles, and graviton-graviton scattering. In this paper, we shall, treating the components as massive scalar point sources, employing the background field method [9,10], and using the scattering amplitude to define the gravitational potential, investigate the quantum corrections to the modified Newtonian potential Eq. (4) in scalarized NS binaries. The paper is organized as follows. Firstly, we will review the effective field theory quantization for gravitation and describe the gravitationally scalarized NS binaries in the effective field theory framework. Then we will give the definition of potential, extract the Feynman rules from the effective action, and look at the calculations of Feymann diagrams. Finally, we summarize our findings and discuss the results.

## 2 Effective field theory description for gravitational scalarized NS binary

Neglecting the extended scales, we treat the star components in scalarized NS binary systems as massive scalar point particles. The action that describes the static[2] scalarized binary reads[3]

$$\begin{aligned} S &= \int d^4x (L_g + L_\phi) \\ &= \int d^4x \sqrt{-g}\left[\frac{2}{\kappa^2}\mathcal{R} - \frac{1}{2}g^{\mu\nu}\partial_\mu\phi\partial_\nu\phi \right.\\ &\quad \left. -\frac{1}{2}m^2\phi^2 + \omega^2\varphi_g^2\phi^2\right]. \end{aligned} \quad (5)$$

Here, $L_\phi$ represents the covariant Lagrangian for scalar configurations settled in the scalarized star components, the quantities $\phi = \phi_{1,2}$, $m = m_{1,2}$, $\omega = \omega_{1,2}$ denote the scalar fields, masses, and scalar charges carried by the components in the system, respectively, and $\varphi_g$ is the gravitational scalar counterpart developed during the dynamical scalarization [28]. $L_g$ and $L_\phi$ represent the effective Lagrangian of the gravitational tensor terms and the terms of massive scalar configurations, respectively. $\kappa = \sqrt{32\pi G}$ is the gravitational coupling. $g = \det g_{\mu\nu}$ denotes the determinant of the gravitational tensor metric $g_{\mu\nu}$. $\mathcal{R} = g^{\mu\nu}\mathcal{R}_{\mu\nu}$ is the Ricci scalar, and the curvature tensor reads,

---

[2] The "static" means that we just consider the velocity-free and separation-dependent gravitational interaction between two stars and ignore the dynamics, which causes orbital decay and thus gravitational radiation from the system.

[3] In the parts for calculating the corrections to gravitational potential, we work with units $c = \hbar = 1$, for simplicity, and choose the metric convention in flat space-time $\eta_{\mu\nu} = \text{diag}(1, -1, -1, -1)$.

$$\mathcal{R}_{\mu\nu} = \partial_\nu \Gamma^\lambda_{\mu\lambda} - \partial_\lambda \Gamma^\lambda_{\mu\nu} + \Gamma^\sigma_{\mu\lambda}\Gamma^\lambda_{\nu\sigma} - \Gamma^\sigma_{\mu\nu}\Gamma^\lambda_{\lambda\sigma}, \quad (6)$$

$$\Gamma^\lambda_{\mu\nu} = \frac{g^{\lambda\sigma}}{2}(\partial_\mu g_{\nu\sigma} + \partial_\nu g_{\mu\sigma} - \partial_\sigma g_{\mu\nu}). \quad (7)$$

In order to treat action (5) as an effective field theory one must include all possible higher derivative couplings of the fields in the gravitational Lagrangian [8]. We consequently write an effective Lagrangian for gravitational tensor terms and scalar configurations in describing the gravitationally scalarized binary as

$$L_g^{\text{eff}} = \frac{2\mathcal{R}}{\kappa^2} + c_1 \mathcal{R}^2 + c_2 \mathcal{R}^{\mu\nu}\mathcal{R}_{\mu\nu} + \cdots, \quad (8)$$

$$\begin{aligned}L_\phi^{\text{eff}} &= -\frac{1}{2} g^{\mu\nu}\partial_\mu\phi\partial_\nu\phi - \frac{1}{2}(m^2 + \omega^2\varphi_g^2)\phi^2 \\ &+ \bar{c}_1 \mathcal{R}^{\mu\nu}\partial_\mu\phi\partial_\nu\phi + \bar{c}_2 \mathcal{R}\partial_\mu\phi\partial^\mu\phi \\ &+ \bar{c}_3 \mathcal{R}(m^2 + \omega^2\varphi_g^2)\phi^2 + \cdots\end{aligned} \quad (9)$$

Here, the coefficients $c_1, c_2, \ldots$ are dimensionless constants that determine the scale of the energy expansion of pure gravity [30], and $\bar{c}_1, \bar{c}_2, \bar{c}_3, \ldots$ are energy-scale dependent coupling constants determined currently by binary observational measurements.

We expand the metric as a background part $\bar{g}_{\mu\nu}$ and a quantum contribution $\kappa h_{\mu\nu}$,

$$\begin{aligned}g_{\mu\nu} &= \bar{g}_{\mu\nu} + \kappa h_{\mu\nu}, \\ g^{\mu\nu} &= \bar{g}^{\mu\nu} - \kappa h^{\mu\nu} + \kappa^2 h^\mu_\lambda h^{\lambda\nu} + \cdots, \\ \sqrt{-g} &= \sqrt{-\bar{g}}\left(1 + \frac{1}{2}\kappa h + \cdots\right),\end{aligned} \quad (10)$$

where $h^{\mu\nu} = \bar{g}^{\mu\alpha}\bar{g}^{\nu\beta}h_{\alpha\beta}$ and $h = \bar{g}^{\mu\nu}h_{\mu\nu}$. For the simplicity of graviton propagator, a gauge fixing term [31,32], with the form of [33]

$$\frac{1}{\kappa^2}\sqrt{-g}\left(\partial_\mu h^{\mu\nu} - \frac{1}{2}\partial^\nu h\right)^2 \quad (11)$$

should be introduced, which then gives the bare graviton propagator

$$\mathcal{D}_{\mu\nu\alpha\beta}(q) = \frac{i\mathcal{P}_{\mu\nu\alpha\beta}}{q^2} = \frac{i(\eta_{\mu\alpha}\eta_{\nu\beta} + \eta_{\mu\beta}\eta_{\nu\alpha} - \eta_{\mu\nu}\eta_{\alpha\beta})}{2q^2}, \quad (12)$$

where $q$ is THE momentum. In calculating quantum corrections at one loop, we need to consider the Lagrangian to quadratic order. Consequently, the explicit expansion of Lagrangian up to the necessary order can be written as follows,

$$L(\phi^2) = -\frac{1}{2}(\partial_\mu\phi\partial^\mu\phi + m^2\phi^2 + \omega^2\varphi_g^2\phi^2), \quad (13)$$

$$\begin{aligned}L(h^2) = -\frac{1}{2}(&h_{\alpha\beta,\gamma}h_{\alpha\beta,\gamma} - 2h_{\gamma\beta,\alpha}h_{\gamma\alpha,\beta} \\ &+ 2h_{\beta\beta,\alpha}h_{\gamma\alpha,\gamma} - h_{\beta\beta,\alpha}h_{\gamma\gamma,\alpha}),\end{aligned} \quad (14)$$

$$\begin{aligned}L(h\phi^2) = \frac{\kappa}{2}\Big[&h_{\mu\nu}\partial_\mu\phi\partial_\nu\phi \\ &- \frac{1}{2}h_{\alpha\alpha}(\partial_\beta\phi\partial_\beta\phi + m^2\phi^2 + \omega^2\varphi_g^2\phi^2)\Big],\end{aligned} \quad (15)$$

$$\begin{aligned}L(h^3) = \frac{\kappa}{2}\bigg[&h_\lambda\left(h_{\gamma\beta,\alpha}h_{\gamma\alpha,\beta} - \frac{1}{2}h_{\alpha\beta,\gamma}h_{\alpha\beta,\gamma}\right. \\ &\left. -h_{\beta\beta,\alpha}h_{\gamma\alpha,\gamma} + \frac{1}{2}h_{\beta\beta,\alpha}h_{\gamma\gamma,\alpha}\right) \\ &+h_{\mu\nu}(h_{\alpha\beta,\mu}h_{\alpha\beta,\nu} - 2h_{\mu\beta,\alpha}h_{\nu\alpha,\beta} \\ &+2h_{\mu\beta,\alpha}h_{\nu\beta,\alpha} + 2h_{\beta\beta,\alpha}h_{\mu\alpha,\nu} \\ &-2h_{\beta\beta,\alpha}h_{\mu\nu,\alpha} + 2h_{\beta\mu,\beta}h_{\alpha\alpha,\nu} \\ &-2h_{\beta\beta,\mu}h_{\alpha\alpha,\nu} + 2h_{\beta\alpha,\beta}h_{\mu\nu,\alpha} - 4h_{\beta\mu,\alpha}h_{\beta\alpha,\nu})\bigg],\end{aligned} \quad (16)$$

$$\begin{aligned}L(h^2\phi^2) = -\frac{\kappa^2}{2}\bigg[&h_{\mu\alpha}h_{\nu\alpha}\partial_\mu\phi\partial_\nu\phi - \frac{1}{2}h_{\alpha\alpha}h_{\mu\nu}\partial_\mu\phi\partial_\nu\phi \\ &+\left(\frac{1}{2}h_{\alpha\alpha}h_{\beta\beta} - \frac{1}{2}h_{\alpha\beta}h_{\alpha\beta}\right)(\partial_\alpha\phi\partial_\alpha\phi + m^2\phi^2 \\ &+\omega^2\varphi_g^2\phi^2)\bigg],\end{aligned} \quad (17)$$

where $h_{\alpha\beta,\gamma} \equiv \frac{\partial h_{\alpha\beta}}{\partial x_\gamma}$.

In some given NS binaries, the gravitational scalar counterparts $\varphi_g$ can become massive [28], with a mass of $m_s$, which may modify the propagator as $\frac{i}{q^2 - m_s^2}$. Accordingly, the gravitational scalar interactions in these systems between two scalarized components are mediated by massive scalar propagator $\frac{i}{q^2 - m_s^2 + i\epsilon}$, which results in analytic contributions to the gravitational potential. Owing to an exponential Yukawa suppression, the propagations of massive mode, with obvious representation as $\frac{1}{q^2 - m_s^2} = -\frac{1}{m_s^2}(1 + \frac{q^2}{m_s^2} + \ldots)$ in momentum $q$, are screened in the range of binary orbit [28]. So the analytical contributions are local effects. The non-analytic effects, arising from the propagations of massless modes of both gravitons and scalar fields, dominate in magnitude over the analytic corrections in the low-energy limit of the effective field theory on large distance. In order to compute the leading long range, low energy quantum corrections to the potential (4), we just consider the non-local, non-analytic contributions to the potential.

## 3 Results from the Feynman diagrams

According to the discussion in Sect. 1 and dimensional analysis, we can figure out the modifications to the potential (4) of the form

$$V(r) = -\frac{Gm_1 m_2}{r} - \frac{G\omega_1 \omega_2}{r} + a_{\text{cl}} \frac{Gm_t}{c^2 r} + a_{\text{qu}} \frac{G\hbar}{c^3 r^2} + \cdots, \quad (18)$$

where $m_t$ contains both gravitational mass ($m_1$, $m_2$) and the scalar contributions ($\omega_1 \varphi_g$, $\omega_2 \varphi_g$). What we shall do in this section is to calculate the numerical efficients $a_{\text{cl}}$ and $a_{\text{qu}}$ for an appropriate definition of potential.

In our calculations, we only consider the non-analytic contributions from the one-loop diagrams, which contain two or more massless propagating particles. The general form for diagrams contributing to the scattering matrix in the momentum ($q$) space representation is

$$\mathcal{M}(q) \sim \left[ A + Bq^2 + \cdots + \alpha \kappa^4 / q^4 + \beta_1 \kappa^4 \ln(-q^2) \right. \\ \left. + \beta_2 \kappa^4 \frac{m}{\sqrt{-q^2}} + \beta_3 \kappa^4 \frac{\omega}{\sqrt{-q^2}} + \cdots \right]. \quad (19)$$

Here, $A$, $B$, ..., in the terms with power series of momentum $q$, correspond to analytic pieces, which only dominate in high-energy regime of the effective field theory and are of no interest to our calculations. The coefficients $\alpha$, $\beta_1$, $\beta_2$, $\beta_3$, ... associate with the long range, non-analytic interactions, where $\beta_1$, $\beta_2$, $\beta_3$ associated terms yield the leading post-Newtonian and quantum corrections to the gravitational potential. The resulting amplitudes are transformed to produce the scattering potential, by performing Fourier transformation and using the following integrals,

$$\int \frac{d^3q}{(2\pi)^3} e^{i\vec{q}\cdot\vec{r}} \frac{1}{|\vec{q}|^2} = \frac{1}{4\pi r},$$
$$\int \frac{d^3q}{(2\pi)^3} e^{i\vec{q}\cdot\vec{r}} \frac{1}{|\vec{q}|} = \frac{1}{2\pi^2 r^2},$$
$$\int \frac{d^3q}{(2\pi)^3} e^{i\vec{q}\cdot\vec{r}} \ln(|\vec{q}|^2) = -\frac{1}{2\pi r^3}. \quad (20)$$

### 3.1 Definition of bound potential

Because the NS binaries are gravitational bound systems, we consider the expectational value for the matrix $iT$ and use the scattering amplitude itself to define the gravitational potential. The full scattering amplitude are calculated in order to represent the non-relativistic potential generated by the non-analytic pieces [15,19],

$$\langle f|iT|i\rangle = (2\pi)^4 \delta^{(4)}(q_1 - q_1' + q_2 - q_2')[i\mathcal{M}(\vec{q})] \\ = -iV(\vec{q})(2\pi)\delta(E - E'). \quad (21)$$

Here, $q_1$, $q_2$ and $q_1'$, $q_2'$ are the incoming and outgoing momentum, respectively. $E - E'$ is the energy difference between the incoming and outgoing states. $\mathcal{M}(\vec{q})$ is the non-analytical part of the amplitude in momentum space representation. $V(\vec{q}) = -\frac{1}{2m_1}\frac{1}{2m_2}\mathcal{M}(\vec{q})$. Taking the non-relativistic limit and Fourier transformation, we can get the corresponding coordinate space representation,

$$V(\vec{r}) = \frac{1}{2m_1}\frac{1}{2m_2}\int \frac{d^3q}{(2\pi)^3} e^{i\vec{q}\cdot\vec{r}} \mathcal{M}(\vec{q}). \quad (22)$$

### 3.2 Vertex rules

From the effective Lagrangian (13–17), our calculations for the Feynman diagrams involve two scalar-one graviton vertex ($h\phi^2$), two scalar-two graviton vertex ($h^2\phi^2$), and three-graviton vertex ($h^3$). The two scalar-one graviton vertex is given by

$$\tau_1^{\mu\nu}(k_1, k_2, m) = \frac{i\kappa}{2}[k_1^\mu k_2^\nu + k_1^\nu k_2^\mu - \eta^{\mu\nu}(k_1 \cdot k_2 - m^2)], \quad (23)$$

where $k_1$, $k_2$ denote the four-momentum of the incoming and outgoing scalar particles, respectively. The two scalar-two graviton vertex can be written as

$$\tau_2^{\mu\nu\alpha\beta}(k_1, k_2, m) = i\kappa^2 \left\{ [I^{\mu\nu\rho\lambda} I_\lambda^{\alpha\beta\sigma} - \frac{1}{4}(\eta_{\mu\nu} I^{\alpha\beta\rho\sigma} + \eta^{\alpha\beta} I^{\mu\nu\rho\sigma})] \right. \\ \times (k_{1\rho} k_{2\sigma} + k_{2\rho} k_{1\sigma}) \\ \left. - \frac{1}{2}\left(I^{\mu\nu\alpha\beta} - \frac{1}{2}\eta^{\mu\nu}\eta^{\alpha\beta}\right) \times [(k_1 \cdot k_2) - m^2] \right\}, \quad (24)$$

where $I_{\mu\nu\alpha\beta} = \frac{1}{2}(\eta_{\mu\alpha}\eta_{\nu\beta} + \eta_{\mu\beta}\eta_{\nu\alpha})$ and the pairs of indices ($\mu\nu$) and ($\alpha\beta$) are associated with two graviton lines. The three-graviton vertex is derived via the background field method, which has the form [8]

$$\tau_{3\alpha\beta\gamma\delta}^{\mu\nu}(k, q) \\ = \frac{i\kappa}{2} \times \left( \mathcal{P}_{\alpha\beta\gamma\delta}\left[k^\mu k^\nu + (k-q)^\mu (k-q)^\nu + q^\mu q^\nu - \frac{3}{2}\eta^{\mu\nu} q^2\right] \right. \\ + 2q_\lambda q_\sigma (I_{\alpha\beta}^{\sigma\lambda} I_{\gamma\delta}^{\mu\nu} + I_{\gamma\delta}^{\sigma\lambda} I_{\alpha\beta}^{\mu\nu} - I_{\alpha\beta}^{\mu\sigma} I_{\gamma\delta}^{\nu\lambda} - I_{\gamma\delta}^{\mu\sigma} I_{\alpha\beta}^{\nu\lambda}) \\ + [q_\lambda q^\mu (\eta_{\alpha\beta} I_{\gamma\delta}^{\nu\lambda} + \eta_{\gamma\delta} I_{\alpha\beta}^{\nu\lambda}) \\ + q_\lambda q^\nu (\eta_{\alpha\beta} I_{\gamma\delta}^{\mu\lambda} - \eta_{\gamma\delta} I_{\alpha\beta}^{\mu\lambda}) - q^2 (\eta_{\alpha\beta} I_{\gamma\delta}^{\mu\nu} - \eta_{\gamma\delta} I_{\alpha\beta}^{\mu\nu}) \\ - \eta^{\mu\nu} q_\sigma q_\lambda (\eta_{\alpha\beta} I_{\gamma\delta}^{\sigma\lambda} - \eta_{\gamma\delta} I_{\alpha\beta}^{\sigma\lambda})] \\ + \{2q_\lambda [I_{\alpha\beta}^{\lambda\sigma} I_{\gamma\delta\sigma}^\nu (k-q)^\mu + I_{\alpha\beta}^{\lambda\sigma} I_{\gamma\delta\sigma}^\mu (k-q)^\nu \\ - I_{\gamma\delta}^{\lambda\sigma} I_{\alpha\beta\sigma}^\nu k^\mu - I_{\gamma\delta}^{\lambda\sigma} I_{\alpha\beta\sigma}^\mu k^\nu] + q^2 (I_{\alpha\beta\sigma}^\mu I_{\gamma\delta}^{\lambda\sigma} + I_{\alpha\beta}^{\nu\sigma} I_{\gamma\delta\sigma}^\mu) \\ + \eta^{\mu\nu} q_\sigma q_\lambda (I_{\alpha\beta}^{\lambda\rho} I_{\gamma\delta\rho}^\sigma + I_{\gamma\delta}^{\lambda\rho} I_{\alpha\beta\rho}^\sigma)\} \\ + \left\{ [k^2 + (k-q)^2][I_{\alpha\beta}^{\mu\sigma} I_{\gamma\delta\sigma}^\nu + I_{\gamma\delta}^{\mu\sigma} I_{\alpha\beta\sigma}^\nu - \frac{1}{2}\eta^{\mu\nu}\mathcal{P}_{\alpha\beta\gamma\delta}] \right. \\ \left. \left. - [I_{\gamma\delta}^{\mu\nu}\eta_{\alpha\beta} k^2 + I_{\alpha\beta}^{\mu\nu}\eta_{\gamma\delta}(k-q)^2] \right\} \right). \quad (25)$$

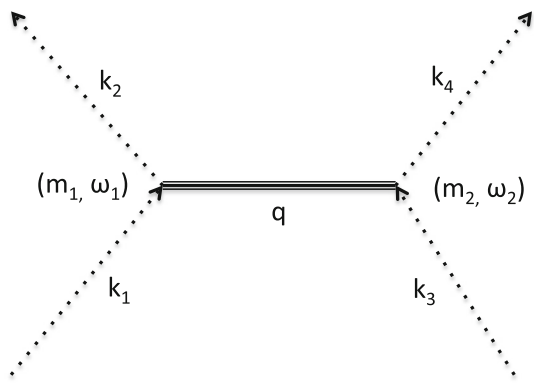

**Fig. 1** The tree diagram contributes to the scalar-modified Newtonian potential. The graviton-graviton scattering yields the Newtonian potential, while the gravitational scalar interaction results from the exchanges of scalar particles, which accompanies the graviton exchanges. The triple solid lines represent the exchanges of gravitons and accompanied exchanges of scalar particles in a scalarized binary system, in which the central thick line denotes the gravitons, and the thin lines on two sides are the scalar particles. The dash lines are the scalar fields

### 3.3 Tree diagram

The set of tree diagrams, coming from the exchanges of both gravitons and scalar particles in Fig. 1, are the well-known lowest order potential in the non-relativistic limit. Because the scalar configuration couples to the star matters inside each component, the exchanges of gravitons actually blend with that of scalar particles. We put them into one single Feynman diagram in Fig. 1. However, the gravitational interactions between star matters realize via the exchanges of gravitons, while the scalar configurations settled in the star components gravitationally interact with each other by the exchanges of scalar particles [27], we separate the exchanges of gravitons from that of scalar particles when calculating the contributions to the potential. By using the Feynman rules and choosing a parameterization of the momentum, the piece of graviton exchanges with a momentum $q$ can be defined as

$$i\mathcal{M}_1^g = \tau_1^{\mu\nu}(k_1,k_2,m_1)\frac{i\mathcal{P}_{\mu\nu\alpha\beta}}{q^2}\tau_1^{\alpha\beta}(k_3,k_4,m_2), \quad (26)$$

where $q = k_1 - k_2 = k_4 - k_3$. The component with mass $m_1$ and scalar charges $\omega_1$ has incoming momentum $k_1$ and outgoing momentum $k_2$, and the other component with mass $m_2$ and scalar charges $\omega_2$ has incoming momentum $k_3$ and outgoing momentum $k_4$, respectively. By contracting all indices for the tree level and performing Fourier transforms, we obtain the scattering potential

$$V_1^g(r) = -\frac{Gm_1m_2}{r}, \quad (27)$$

which gives the Newtonian law.

The settled scalar configurations inside the components enhance stars' masses [26] and subsequently the gravita-tional attraction, which makes the scalarized system act as a source of emission of the gravitational scalar counterpart [28]. Therefore, the vertex associated with the propagations of scalar fields between two stars involves scalar mass dimensional quantities $\omega_1\varphi_g$ and $\omega_2\varphi_g$, depending on the strength of gravitational scalar counterpart $\varphi_g$. The scalar configuration, i.e. the spontaneous scalarization of NS, are the byproduct for a more compact NS, with a compactness above a certain value. As a consequence, the exchanges of scalar particles between two scalarized star components must accompany the exchanges of gravitons, which is given express to the vertex. The appearance of the gravitational scalar counterpart in the scalar mass of the star components just affect the strength of the scalar charges, by means of the coupling strength with the star material, and indirectly change the star masses, which doesn't play the direct role in the exchanges of the scalar particles in scattering process. Accordingly, we define the scattering of scalar particles with a propagating momentum $l$ as

$$i\mathcal{M}_1^s = \tau_1^{\mu\nu}(k_1,l,\omega_1)\frac{i}{l^2}\tau_1^{\alpha\beta}(l,k_3,\omega_2). \quad (28)$$

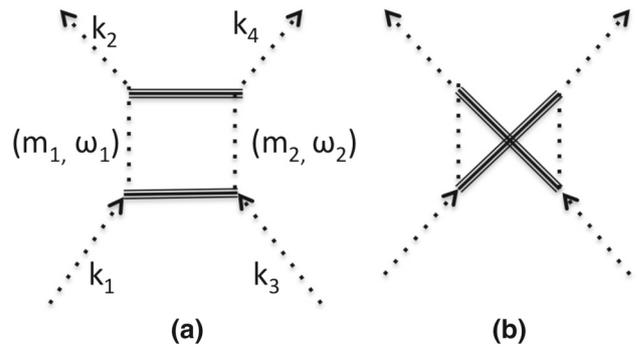

**Fig. 2** The set of box (**a**) and crossed-box (**b**) diagrams contribute to the non-analytic components of the potential

In non-relativistic limit, we perform Fourier transformation

$$V(\vec{r}) = \frac{1}{2\omega_1}\frac{1}{2\omega_2}\int\frac{d^3q}{(2\pi)^3}e^{i\vec{q}\cdot\vec{r}}\mathcal{M}(\vec{q}), \quad (29)$$

and find the result

$$V_1^s(r) = -\frac{G\omega_1\omega_2}{r}, \quad (30)$$

which arises from the scalar-modified classical Newtonian gravitational interaction between two scalarized stars. The combined result of two pieces yields the modified Newtonian potential (4).

### 3.4 Box and crossed-box diagrams

The box (Fig. 2a) and crossed-box (Fig. 2b) diagrams just involve vertex ($h\phi^2$). Each diagram contains the exchanges both of gravitons and of scalar particles.

**Fig. 3** The set of triangle diagrams contribute to the non-analytic pieces of potential

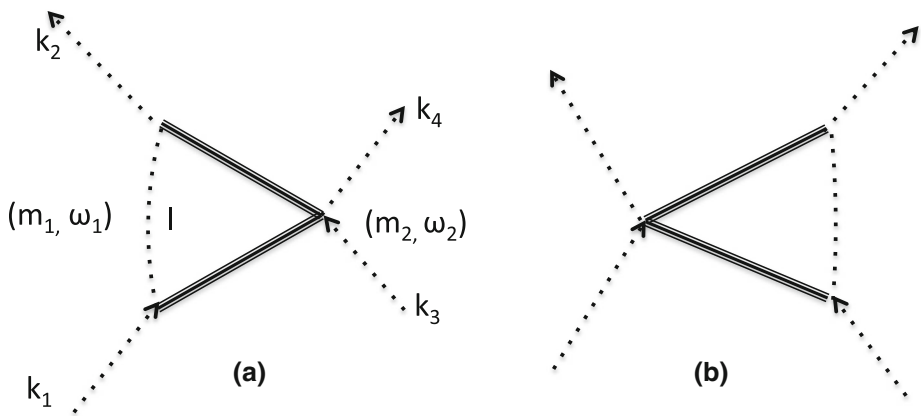

In the box diagram Fig. 2a, the contributions are written as

$$i\mathcal{M}^g_{2(a)} = \int \frac{d^4l}{(2\pi)^4} \frac{i}{(k_1+l)^2 - m_1^2} \frac{i\mathcal{P}_{\mu\nu\alpha\beta}}{l^2} \frac{i\mathcal{P}_{\rho\sigma\gamma\delta}}{(l+q)^2} \frac{i}{(k_3-l)^2 - m_2^2}$$
$$\times \tau_1^{\mu\nu}(k_1, k_1+l, m_1) \tau_1^{\rho\sigma}(k_1+l, k_2, m_1)$$
$$\times \tau_1^{\alpha\beta}(k_3, k_3-l, m_2) \tau_1^{\gamma\delta}(k_3-l, k_4, m_2), \quad (31)$$

for gravitational tensor interactions realized by the exchanges of gravitons, and as

$$i\mathcal{M}^s_{2(a)}$$
$$= \int \frac{d^4l}{(2\pi)^4} \frac{i}{(l+k_1)^2 - m_1^2} \frac{i}{(l-k_3)^2 - m_2^2} \frac{i}{l^2} \frac{i}{(l+q')^2}$$
$$\tau_1^{\mu\nu}(k_1, l, \omega_1) \tau_1^{\rho\sigma}(k_2, l+q', \omega_1)$$
$$\tau_1^{\alpha\beta}(l, k_3, \omega_2) \tau_1^{\gamma\delta}(k_3, l+q', \omega_2) \quad (32)$$

for the gravitational scalar interactions via the exchanges of scalar particles, which must associate with the exchanges of gravitons. The non-analytic contributions to the potential from the crossed-box in Fig. 2b are

$$i\mathcal{M}^g_{2(b)}$$
$$= \int \frac{d^4l}{(2\pi)^4} \frac{i}{(k_1+l)^2 - m_1^2} \frac{i\mathcal{P}_{\mu\nu\alpha\beta}}{l^2} \frac{i\mathcal{P}_{\rho\sigma\gamma\delta}}{(l+q)^2} \frac{i}{(k_4+l)^2 - m_2^2}$$
$$\times \tau_1^{\mu\nu}(k_1, k_1+l, m_1) \tau_1^{\rho\sigma}(k_1+l, k_2, m_1)$$
$$\tau_1^{\alpha\beta}(k_3, k_4+l, m_2) \tau_1^{\gamma\delta}(k_4+l, k_4, m_2) \quad (33)$$

for graviton exchanges and

$$i\mathcal{M}^s_{2(b)}$$
$$= \int \frac{d^4l}{(2\pi)^4} \frac{i}{(k_1+l)^2 - m_1^2} \frac{i}{(k_4+l)^2 - m_2^2} \frac{i}{(l-q')^2} \frac{i}{l^2}$$
$$\tau_1^{\mu\nu}(k_1, l, \omega_1) \tau_1^{\rho\sigma}(k_1, l-q', \omega_1)$$
$$\tau_1^{\alpha\beta}(l-q', k_3, \omega_2) \tau_1^{\gamma\delta}(k_4, l, \omega_2) \quad (34)$$

for exchanges of scalar particles. The star with mass and scalar charge $(m_1, \omega_1)$ has incoming momentum $k_1$ and outgoing momentum $k_2$, and the other component with $(m_2, \omega_2)$ has incoming momentum $k_3$ and outgoing momentum $k_4$, respectively.

For the calculations of diagrams, we employ the algebraic program and the contraction rules, which are discussed in references [19,34] in order to reduce the integrals, and we also use the integrals listed in the appendix in these two references. The results from graviton exchanges are in agreement with that of [19], i.e.

$$V^g_{2(a)+2(b)}(r) = -\frac{47}{3\pi} \frac{Gm_1 m_2}{r} \frac{G}{r^2}. \quad (35)$$

The contributions from exchanges of scalar particles give

$$V^s_{2(a)+2(b)}(r) = -\frac{G\omega_1\omega_2}{r}\left(\frac{2G(m_1+m_2)}{3r} + \frac{11}{3\pi}\frac{G}{r^2}\right). \quad (36)$$

### 3.5 Triangle diagrams

The triangle diagrams contributing to the non-analytic pieces arise from graviton exchanges involving in the effective Lagrangian $L(h\phi^2)$ and $L(h^2\phi^2)$ (see Fig. 3). The expressions are same as [19],

$$i\mathcal{M}^g_{3(a)} = \int \frac{d^4l}{(2\pi)^4} \frac{i}{(k_1+l)^2 - m_1^2} \frac{i\mathcal{P}_{\mu\nu\alpha\beta}}{l^2} \frac{i\mathcal{P}_{\rho\sigma\gamma\delta}}{(l+q)^2}$$
$$\tau_1^{\mu\nu}(k_1, k_1+l, m_1) \tau_1^{\rho\sigma}(k_1+l, k_2, m_1)$$
$$\tau_2^{\alpha\beta\gamma\delta}(k_3, k_4, m_2), \quad (37)$$

$$i\mathcal{M}^g_{3(b)} = \int \frac{d^4l}{(2\pi)^4} \frac{i\mathcal{P}_{\mu\nu\alpha\beta}}{l^2} \frac{i\mathcal{P}_{\rho\sigma\gamma\delta}}{(l+q)^2} \frac{i}{(l-k_3)^2 - m_2^2}$$
$$\tau_1^{\alpha\beta}(k_3, k_3-l, m_2) \tau_1^{\gamma\delta}(k_3-l, k_4, m_2)$$
$$\tau_2^{\mu\nu\rho\sigma}(k_1, k_2, m_1). \quad (38)$$

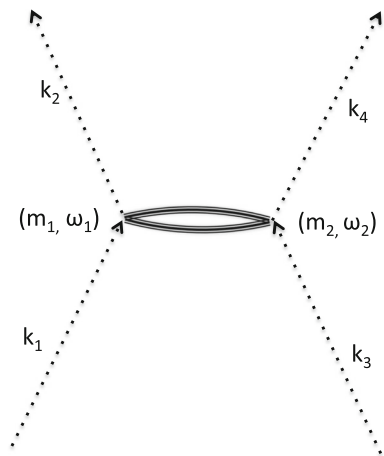

**Fig. 4** The circular diagram contribute to the non-analytic components of the potential

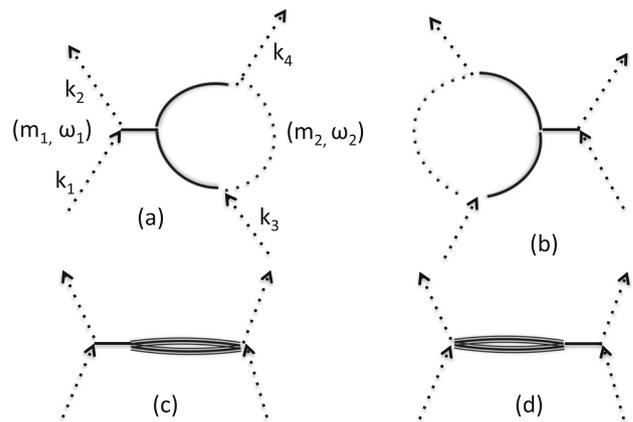

**Fig. 5** Two classes of the set of one particle reducible diagrams yield non-analytic corrections to the potential. The solid lines denote the gravitons

Taking the non-relativistic limit, we reproduce the results

$$V^g_{3(a)+3(b)} = -\frac{Gm_1m_2}{r}\left[4\frac{G(m_1+m_2)}{r} - \frac{28}{\pi}\frac{G}{r^2}\right]. \quad (39)$$

The pieces from the exchange of scalar particles have the following expressions,

$$i\mathcal{M}^s_{3(a)} = \int \frac{d^4l}{(2\pi)^4} \frac{i}{(k_1+l)^2 - m_1^2} \frac{i}{l^2} \frac{i}{(l+q)^2}$$
$$\tau_1^{\mu\nu}(k_1,l,\omega_1)\tau_1^{\rho\sigma}(l,k_2,\omega_1)\tau_2^{\alpha\beta\gamma\delta}(l,l+q,\omega_2), \quad (40)$$

$$i\mathcal{M}^s_{3(b)} = \int \frac{d^4l}{(2\pi)^4} \frac{i}{l^2} \frac{i}{(l+q)^2} \frac{i}{(k_3+l)^2-m_2^2}$$
$$\tau_2^{\mu\nu\rho\sigma}(l,l+q,\omega_1)\tau_1^{\alpha\beta}(l,k_3,\omega_2)\tau_1^{\gamma\delta}(l+q,k_3,\omega_1). \quad (41)$$

The non-analytic contributions to the potential is obtained as

$$V^s_{3(a)+3(b)} = -\frac{G\omega_1\omega_2}{r}\left[\frac{4G(\omega_1+\omega_2)}{r} + \frac{6}{\pi}\frac{G}{r^2}\right]. \quad (42)$$

### 3.6 Circular diagram

By taking the symmetry into account, we write down the expressions of circular diagrams involving in both gravitons exchange (Fig. 4a) and the exchange of scalar particles (Fig. 4b) as follows, respectively,

$$i\mathcal{M}^g_4 = \frac{1}{2!}\int \frac{d^4l}{(2\pi)^4}\tau_2^{\mu\nu\rho\sigma}(k_1,k_2,m_1)\tau_2^{\alpha\beta\gamma\delta}$$
$$\times(k_3,k_4,m_2)\frac{i\mathcal{P}_{\mu\nu\alpha\beta}}{(l+q)^2}\frac{i\mathcal{P}_{\rho\sigma\gamma\delta}}{l^2}, \quad (43)$$

$$i\mathcal{M}^s_4 = \frac{1}{2!}\int \frac{d^4l}{(2\pi)^4}\tau_2^{\mu\nu\rho\sigma}(k_1,k_2,\omega_1)\tau_2^{\alpha\beta\gamma\delta}$$
$$\times(k_3,k_4,\omega_2)\frac{i}{(l+q)^2}\frac{i}{l^2}. \quad (44)$$

Performing the same contractions and integrals as reference [19], we obtain the corrections to the potential,

$$V^g_4(r) = -\frac{22}{\pi}\frac{Gm_1m_2}{r}\frac{G}{r^2}, \quad (45)$$

$$V^s_4(r) = \frac{2}{\pi}\frac{G\omega_1\omega_2}{r}\frac{G}{r^2}. \quad (46)$$

### 3.7 One particle reducible diagrams

There are two classes of set of one particle reducible (1PR) diagrams (see Fig. 5). One class of set are the massive loop diagrams (Fig. 5a, b), whose expressions contributing to non-analytic corrections, can be written as,

$$i\mathcal{M}_{5(a)} = \int \frac{d^4l}{(2\pi)^4} \frac{i\mathcal{P}_{\mu\nu\rho\sigma}}{q^2}\frac{i\mathcal{P}_{\lambda\kappa\alpha\beta}}{l^2}\frac{i\mathcal{P}_{\phi\epsilon\gamma\delta}}{(l+q)^2}\frac{i}{(l-k_3)^2-m_2^2}$$
$$\times \tau_1^{\mu\nu}(k_1,k_2,m_1)\tau_3^{\rho\sigma\lambda\kappa(\phi\epsilon)}(-l,q)$$
$$\tau_1^{\alpha\beta}(k_3,k_3-l,m_2)\tau_1^{\gamma\delta}(k_3-l,k_4,m_2), \quad (47)$$

$$i\mathcal{M}_{5(b)} = \int \frac{d^4l}{(2\pi)^4}\frac{i}{(l+k_1)^2-m_1^2}\frac{i\mathcal{P}_{\mu\nu\phi\epsilon}}{l^2}\frac{i\mathcal{P}_{\rho\sigma\lambda\kappa}}{(l+q)^2}\frac{i\mathcal{P}_{\gamma\delta\alpha\beta}}{q^2}$$
$$\times \tau_1^{\mu\nu}(k_1,l+k_1,m_1)\tau_1^{\rho\sigma}(l+k_1,k_2,m_1)$$
$$\tau_3^{\phi\epsilon\gamma\delta(\lambda\kappa)}(-l,q)\tau_1^{\alpha\beta}(k_3,k_4,m_2), \quad (48)$$

which yield the contributions to the potential,

$$V_{5(a)+5(b)} = -\frac{Gm_1m_2}{r}\left[-\frac{G(m_1+m_2)}{r} + \frac{5}{3\pi}\frac{G}{r^2}\right]. \quad (49)$$

The other class of set contains diagrams involving both pure graviton exchanges and the incident mixed scalar-graviton exchanges (Fig. 5c, d). It is represent in the loop,

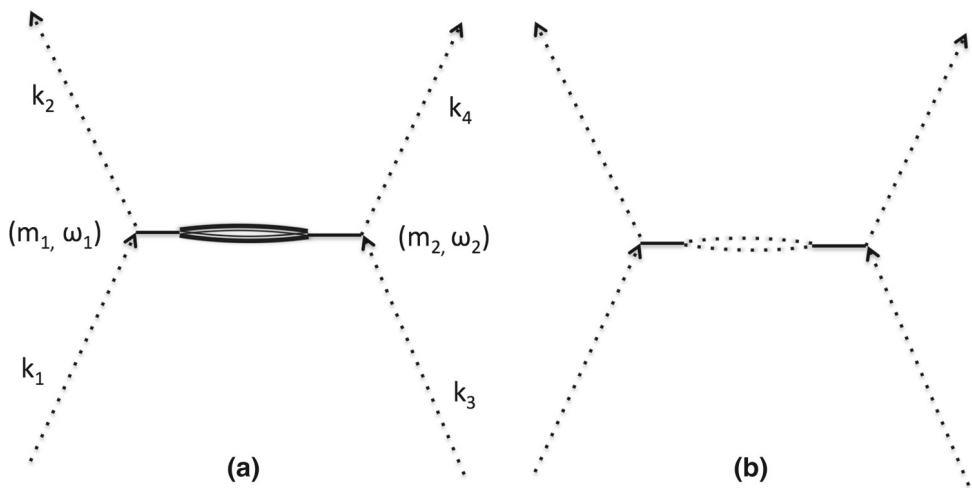

**Fig. 6** The set of vacuum polarization diagrams contribute to the non-analytic corrections to the potential. The graviton loop diagram contains a ghost one, which is marked by the double line in (**a**)

which denotes the exchanges of scalar particles, for the incident mixed scalar-graviton exchanges. By noting that a symmetry factor of $1/2!$, the pure graviton exchanges can be defined as follows,

$$i\mathcal{M}^g_{5(c)} = \frac{1}{2!} \int \frac{d^4l}{(2\pi)^4} \frac{i\mathcal{P}_{\mu\nu\phi\epsilon}}{q^2} \frac{i\mathcal{P}_{\rho\sigma\alpha\beta}}{l^2} \frac{i\mathcal{P}_{\phi\epsilon\gamma\delta}}{(l+q)^2}$$
$$\times \tau_1^{\mu\nu}(k_1, k_2, m_1) \tau_3^{\rho\sigma\lambda\kappa(\phi\epsilon)}(l, -q)$$
$$\tau_2^{\alpha\beta\gamma\delta}(k_3, k_4, m_2), \quad (50)$$

$$i\mathcal{M}^g_{5(d)} = \frac{1}{2!} \int \frac{d^4l}{(2\pi)^4} \frac{i\mathcal{P}_{\mu\nu\phi\epsilon}}{l^2} \frac{i\mathcal{P}_{\rho\sigma\lambda\kappa}}{(l+q)^2} \frac{i\mathcal{P}_{\gamma\delta\alpha\beta}}{q^2}$$
$$\times \tau_2^{\mu\nu\rho\sigma}(k_1, k_2, m_1) \tau_3^{\gamma\delta\phi\epsilon(\lambda\kappa)}(-l, q)$$
$$\tau_1^{\alpha\beta}(k_3, k_4, m_2), \quad (51)$$

and the contributions from the mixed scalar-graviton diagrams are defined as

$$i\mathcal{M}^m_{5(c)} = \frac{1}{2!} \int \frac{d^4l}{(2\pi)^4} \frac{i\mathcal{P}_{\mu\nu\rho\sigma}}{q^2} \frac{i}{l^2} \frac{i}{(l+q)^2}$$
$$\times \tau_1^{\mu\nu}(k_1, k_2, m_1) \tau_1^{\rho\sigma}(l, l+q)$$
$$\tau_2^{\gamma\delta\alpha\beta}(k_3, l+q, \omega_2), \quad (52)$$

$$i\mathcal{M}^m_{5(d)} = \frac{1}{2!} \int \frac{d^4l}{(2\pi)^4} \frac{i}{l^2} \frac{i}{(l+q)^2} \frac{i\mathcal{P}_{\gamma\delta\alpha\beta}}{q^2}$$
$$\times \tau_2^{\mu\nu\rho\sigma}(k_1, -l-q, \omega_1) \tau_1^{\gamma\delta}(-l, -l-q)$$
$$\tau_1^{\alpha\beta}(k_3, k_4, m_2). \quad (53)$$

By performing the algebra analysis, we obtain the following corrections to the potential, respectively,

$$V^g_{5(c)+5(d)} = \frac{26}{3\pi} \frac{Gm_1m_2}{r} \frac{G}{r^2}, \quad (54)$$

$$V^m_{5(c)+5(d)} = -\frac{1}{6\pi^2} \frac{G(\omega_1^2 m_2^2 + \omega_2^2 m_1^2)}{m_1 m_2} \frac{G}{r^3}. \quad (55)$$

### 3.8 Vacuum polarization diagrams

By considering the gauge choice Eq. (11), which yields a Faddeev-Popov ghost along with the graviton loop, we more intuitively separate the graviton loop from scalar loop for the vacuum polarization diagrams. In the graviton loop diagram in Fig. 6a, a ghost loop exists. Accordingly, the vacuum polarization Fig. 6a for graviton and ghost loop diagram has the expression

$$i\mathcal{M}_{6(a)} = \tau_{\mu\nu}(k_1, k_2, m_1) \frac{i\mathcal{P}^{\mu\nu\rho\sigma}}{q^2}$$
$$\Pi_{\rho\sigma\gamma\delta} \frac{i\mathcal{P}^{\gamma\delta\alpha\beta}}{q^2} \tau_{\alpha\beta}(k_3, k_4, m_2), \quad (56)$$

where the vacuum polarization tensor $\Pi_{\rho\sigma\gamma\delta}$ [35,36] satisfies the Slavnov-Taylor identity $q_\mu q_\nu D_{\mu\nu\rho\sigma}(q)\Pi_{\rho\sigma\gamma\delta}(q) D_{\gamma\delta\alpha\beta}(q) = 0$. It gives the contributions to the potential

$$V_{6(a)} = -\frac{43}{30\pi} \frac{Gm_1m_2}{r} \frac{G}{r^2}. \quad (57)$$

While the corrections from the scalar loop diagram Fig. 6b, which involves in a symmetry factor of $1/2!$, can be written as

$$i\mathcal{M}_{6(b)} = \frac{1}{2!} \int \frac{d^4l}{(2\pi)^4} \frac{i\mathcal{P}_{\mu\nu\rho\sigma}}{q^2} \frac{i\mathcal{I}_{\rho\sigma\gamma\delta}}{l^2} \frac{i\mathcal{I}_{\rho\sigma\gamma\delta}}{(l+q)^2} \frac{i\mathcal{P}_{\gamma\delta\alpha\beta}}{q^2}$$
$$\times \tau_1^{\mu\nu}(k_1, k_2, m_1) \tau_1^{\rho\sigma}(l, l+q)$$
$$\tau_1^{\gamma\delta}(-l, -l-q) \tau_1^{\alpha\beta}(k_3, k_4, m_2). \quad (58)$$

We find the results from scalar loop vacuum polarization diagram,

$$V_{6(b)} = -\frac{1}{20\pi} \frac{Gm_1m_2}{r} \frac{G}{r^2}. \quad (59)$$

## 4 Summary and discussions

Adding up all the non-analytical contributions, we get the final corrected gravitational potential

$$V(r) = -\frac{Gm_1m_2}{r}\left[1 + \frac{3G(m_1+m_2)}{c^2 r} + \frac{81}{20\pi}\frac{G\hbar}{c^3 r^2}\right]$$
$$-\frac{G\omega_1\omega_2}{r}\left\{1 + \left[\frac{2(m_1+m_2)}{3} + 4(\omega_1+\omega_2)\right]\frac{G}{c^2 r}\right.$$
$$\left.+\left(\frac{23}{3\pi} + \frac{\frac{\omega_1}{\omega_2}\frac{m_2}{m_1} + \frac{\omega_2}{\omega_1}\frac{m_1}{m_2}}{6\pi^2}\right)\frac{G\hbar}{c^3 r^2}\right\}. \quad (60)$$

It can be found that the gravitational interactions involving exchanges of either gravitons or scalar particles contribute to both classical relativistic post-Newtonian corrections and quantum corrections. In order to look at the corrections clearly, we rewrite the corrected potential (60) as follows,

$$V(r) = -\frac{Gm_1m_2}{r} - \frac{G\omega_1\omega_2}{r} - \frac{3G^2 m_1 m_2(m_1+m_2)}{c^2 r^2}$$
$$-\frac{2G^2 \omega_1\omega_2(m_1+m_2)}{3c^2 r^2} - \frac{4G^2 \omega_1\omega_2(\omega_1+\omega_2)}{c^2 r^2}$$
$$-\frac{81}{20\pi}\frac{Gm_1m_2}{r}\frac{G\hbar}{c^3 r^2} - \frac{23}{3\pi}\frac{G\omega_1\omega_2}{r}\frac{G\hbar}{c^3 r^2}$$
$$-\frac{G(\omega_1^2 m_2^2 + \omega_2^2 m_1^2)}{6\pi^2 m_1 m_2 r}\frac{G\hbar}{c^3 r^2}. \quad (61)$$

We can see different types of terms in the rewritten expression (61). The first two terms are modified Newtonian potential for a scalarized binary system, which represent the lowest order interactions of the two stars and dominate the potential at low energies. The next three terms denote the classical relativistic corrections to the gravitational potential, which are the leading post-Newtonian corrections in general relativity with the scalar charged stars. The classical relativistic corrections arise from just pure particle exchanges. The pure graviton exchanges contribute to the corrections to Newtonian piece, while the pure scalar exchanges produce the contributions to the scalar modified part of the potential. The last three terms represent the leading 1-loop corrections to gravitational potential of scalarized binaries from a quantum point of view. Looking at the quantum corrections, we notice that the contributions arising from graviton exchanges combine into one term, while that resulting from the exchanges involving scalar particles are split up into two terms. There is one term where the two scalar charges are multiplied together, which comes from the combined contributions of pure scalar exchanges. The last term originates from the mixed graviton-scalar particle exchanges in 1PR diagrams, in which the exchanges of scalar particles in the loop always accompany that of gravitons, which are represent by triple lines in diagrams Fig. 5c, d. The two scalar charges are squared and separated in the mixed graviton-scalar particle exchanges, because of the different binding energy of two star components, and the dependence of the scalar charges on the stars' density [29], even in double NS systems.

By considering the dependence of scalar charges on the stars' masses, the scalar charges, which is parametrized by the sensitivities ($s = 0.2$ for NSs and $s = 10^{-4}$ for white dwarfs), range from $10^{-4} - 0.1$ [27]. Therefore, the gravitational scalar effect of the potential is $\frac{G\omega_1\omega_2}{r} \sim 10^{-18} - 10^{-13}$ m/r in scalarized double NS systems, which is more smaller in NS binaries with white dwarf or main sequence companion stars. In the SI units, we can estimate the effect of quantum corrections, i.e. $\frac{G\hbar}{c^3} \sim 10^{-70}$m$^2$, which is indeed small and seemingly impossible to be detected by current astronomical observations.

**Acknowledgements** This work is supported by the Fundamental Research Funds for the Central Universities (Grant no. 161gpy49) at Sun Yat-sen University, the Science and Technology Program of Guangzhou (Grant no. 71000-42050001).

**Data Availability Statement** This manuscript has no associated data or the data will not be deposited. [Authors' comment: This is a theoretical study and no experimental data has been listed.]